\documentclass[aps,prl,twocolumn,groupedaddress,showpacs]{revtex4}
\usepackage{graphicx}
\usepackage{amsmath,amssymb,amsfonts}
\usepackage{natbib}

\newcommand{\be}{\begin{equation}}
\newcommand{\ee}{\end{equation}}

\begin{document}

\title{Line Shape Broadening in Surface Diffusion of Interacting
Adsorbates\\ with Quasielastic He Atom Scattering}

\author{R. Mart\'{\i}nez--Casado$^{a,c}$}
\email{ruth@imaff.cfmac.csic.es}

\author{J.L. Vega$^{b,c}$}
\email{jlvega@imaff.cfmac.csic.es}

\author{A.S. Sanz$^c$}
\email{asanz@imaff.cfmac.csic.es}

\author{S. Miret--Art\'es$^c$}
\email{s.miret@imaff.cfmac.csic.es}

\affiliation{$^a$Lehrstuhl f\"ur Physikalische Chemie I,
Ruhr--Universit\"at Bochum, D--44801 Bochum, Germany;}

\affiliation{$^b$Biosystems Group, School of Computing, University of Leeds,
Leeds LS2 9JT, United Kingdom;}

\affiliation{$^c$Instituto de Matem\'aticas y F\'{\i}sica Fundamental,
Consejo Superior de Investigaciones Cient\'{\i}ficas,
Serrano 123, 28006 Madrid, Spain}

\date{\today}

\begin{abstract}
The experimental line shape broadening observed in adsorbate diffusion
on metal surfaces with increasing coverage is usually related to the
nature of the adsorbate-adsorbate interaction.
Here we show that this broadening can also be understood in terms of
a fully stochastic model just considering two noise sources: (i)
a Gaussian white noise accounting for the surface friction, and
(ii) a shot noise replacing the physical adsorbate-adsorbate
interaction potential.
Furthermore, contrary to what could be expected, for relatively weak
adsorbate-substrate interactions the opposite effect is predicted:
line shapes get narrower with increasing coverage.
\end{abstract}


%68.35.Fx Diffusion; interface formation
%05.10.Gg Stochastic analysis methods (Fokker-Planck, Langevin, etc.)
%68.43.Jk Diffusion of adsorbates, kinetics of coarsening and aggregation

\maketitle

%%%%%%%%%%%%%%%%%%%%%%%%%%%%%%%%%%%%%%%%%%%%%%%%%%%%%%%%%%%%%%%%%%%%%%%

The diffusion of atoms, molecules, or small clusters on metal surfaces
is an elementary dynamical process of paramount importance; it
constitutes the ground step to understanding more complex phenomena
in surface science.
Processes such as heterogeneous catalysis, molecular-beam epitaxy,
crystal growth, chemical vapor deposition, or associative desorption
are all strongly affected by the kinetics of diffusion.
Different experimental techniques have been used to study activated
surface diffusion \cite{gomer,eli,jardine1}, quasielastic helium-atom
scattering (QHAS) being one of the most valuable for such a purpose
\cite{toennies1,toennies2,toennies3}.
In QHAS experiments the observable is the so-called
{\it differential reflection coefficient}.
In analogy to neutron scattering by liquids \cite{vanHove}, this
magnitude gives the probability that the He-atom beam scattered from
the diffusing collective reaches a solid angle $\Omega$ with an energy
exchange $\hbar\omega$=$E_f - E_i$ and parallel (to the surface)
momentum transfer $\Delta {\bf K}$=${\bf K}_f - {\bf K}_i$, and is
proportional to the {\it dynamic structure factor} or {\it scattering
law}, $S(\Delta {\bf K},\omega)$.
$S(\Delta {\bf K},\omega)$ provides information about the dynamics and
structure of the adsorbates, thus allowing to better understand the
nature of adsorbate-substrate \cite{toennies2} and adsorbate-adsorbate
\cite{toennies3} interactions.
In general, $S(\Delta {\bf K},\omega)$ usually consists of the
{\it quasielastic} ($Q$) {\it peak} and some weaker peaks attributed
to the diffusion process and to phonon excitations and adsorbate low
frequency vibrational modes [namely, the {\it frustrated translational}
($T$) {\it modes}], respectively.

Here we present a novel and insightful stochastic approach that allows
to study and understand in a simple manner how the adsorbate coverage
influences $S(\Delta {\bf K},\omega)$.
This approach, grounded on the theory of spectral-line collisional
broadening developed by Van Vleck and Weisskopf \cite{weiss} and the
elementary kinetic theory of gases \cite{mcquarrie} in the Langevin
framework, relies on three assumptions:
(i) the adsorbate-substrate interaction is described by
a (deterministic) adiabatic adsorption potential, (ii) the temperature
dependent nonadiabatic coupling to the substrate electronic and
vibrational excitations is accounted for by a Gaussian white noise,
and (iii) adsorbate-adsorbate interactions are described by a shot
noise \cite{gardiner,hanggi2}, as proposed in Ref.~\onlinecite{pre-I}.
We call this approach the {\it interacting single adsorbate approximation}.
Gaussian white noise has been used to characterize the rich variety of
different transport regimes in surface diffusion \cite{sancho}.
Within a similar context, the shot noise has been used to study thermal
ratchets \cite{hanggi1}, mean first passage times \cite{iturbe}, and
jump distributions \cite{tommei}.

To explain the line-shape broadening undergone by the $Q$ peak with
increasing coverage ($\theta$), high demanding molecular dynamics
calculations within the Langevin framework have been carried out
\cite{toennies3,ying}.
Though the calculations could reproduce the trend with $\theta$,
the dipole-dipole interaction potential considered was not able to
correctly provide the experimental broadening observed.
As we show here, this $\theta$-dependent behavior can also be
reproduced by a simple collision model, with collisions being
described by a shot noise.
Note that, providing collective behaviors are not relevant (as happens,
for instance, at low temperatures \cite{benedek}), adsorbates undergo
a relatively large number of collisions for relatively long time scales
reaching the {\it statistical} or {\it stochastic limit}.
This proves that the nature of the adsorbate-adsorbate interaction could
not be very relevant regarding the line shape broadening (as well as its
trend).
As seen here, the broadening is strikingly related to the surface
corrugation (which strongly couples diffusion and low frequency
vibrations) as well as to the friction associated to increasing
$\theta$.
This important result emerges when combining the numerical simulations
using our model with a fully analytical treatment, which allows to
understand and to interpret the process in terms of two limiting cases:
{\it running or diffusive} and {\it bound trajectories} \cite{risken}.

The starting point of our approach consists of expressing the dynamic
structure factor as \cite{vanHove,JLvega1}
\be
 S(\Delta {\bf K},\omega) =
  \int e^{-i\omega t} \ \! I(\Delta{\bf K},t) dt ,
 \label{eq2}
\ee
where
\be
 I(\Delta {\bf K},t) = \langle e^{-i\Delta {\bf K} \cdot
   [{\bf R}(t) - {\bf R}(0)] } \rangle
  = \langle e^{-i \Delta K \!
   \int_0^t v_{\Delta {\bf K}} (t') dt'} \rangle
 \label{eq3}
\ee
is the {\it intermediate scattering function}.
In (\ref{eq3}), the brackets denote an ensemble average and
$v_{\Delta {\bf K}}$ is the adparticle velocity projected onto the
$\Delta {\bf K}$ direction ($\Delta K$=$\| \Delta {\bf K} \|$).
After a second order cumulant expansion in $\Delta K$ in the right-hand
side (rhs) of the second equality of (\ref{eq3}), one obtains
\begin{equation}
 I(\Delta {\bf K},t) \approx
  e^{- \Delta K^2 \! \int_0^t (t - t') \mathcal{C}(t') dt'} ,
 \label{eq4}
\end{equation}
where $\mathcal{C}(t)$ =
$\langle v_{\Delta {\bf K}}(0) v_{\Delta {\bf K}}(t) \rangle$ is the
{\it velocity autocorrelation function}.
This is the so-called {\it Gaussian approximation} \cite{mcquarrie},
which is exact when velocity correlations of third or higher order are
negligible.
Despite its limitations, it provides much insight into the dynamical
process by allowing an analytical treatment of the problem.

Determining the line shape of $S(\Delta {\bf K},\omega)$ through
(\ref{eq2}) thus requires us to simulate the adsorbate dynamics.
This is done within a purely Langevin framework that includes the
three elements mentioned above.
Considering one dimension for simplicity, the motion of an adsorbate
interacting with another adsorbates and a temperature-dependent
substrate is ruled by the generalized Langevin equation \cite{hanggi2}
\begin{equation}
 \ddot{x}(t) = - \int_0^t \eta (t-t') \ \! \dot{x}(t') \ \! dt'
  + F[x(t)] + \delta R(t) ,
 \label{eq5}
\end{equation}
where $x$ is the adsorbate coordinate, $\eta(t)$ is the memory
function, $F$=$- \nabla V$ is the deterministic force per mass unit
($V$ is the periodic adsorbate-surface interaction potential, with
period $a$), and $\delta R(t)$=$\delta R_G (t)$+$\delta R_S (t)$ is the
noise source ($G$ and $R$ refer to Gaussian white noise and shot noise,
respectively).

Gaussian white noise is defined by $\langle R_G (t) \rangle$=0 and
$\langle R_G(t) R_G(t') \rangle$=$2m\gamma k_B T \delta(t'-t)$,
where $m$ is the adsorbate mass, $T$ is the surface temperature, and
$\gamma$ is the (constant) friction coefficient measuring the strength
of the adsorbate-substrate coupling.
On the other hand, the shot noise term in our model is
$R_S (t)$=$\sum_{k=1}^K b_k (t - t_k)$, where $b_k(t-t_k)$=$c_k \lambda'
{\rm e}^{- \lambda' (t-t_k)}$ is the pulse or impact shape, with
$t-t_k$$>$0 and $c_k$ giving the intensity of the collision impact.
The pulse shape indicates that collisions are assumed to be sudden
(strong but elastic) and after-collision effects relax exponentially
at a constant rate $\lambda'$.
The probability for $K$ collisions to happen after a time $\mathcal{T}$
follows a Poisson distribution \cite{gardiner},
$P_K (\mathcal{T})$=$(K!)^{-1} (\lambda \mathcal{T})^K
e^{-\lambda\mathcal{T}}$, where $\lambda$ means the average number of
collisions per time unit.
Since collisions take place randomly at different orientations and
energies, we reasonably assume $g(c_k)$=$\alpha^{-1} e^{-c_k/\alpha}$,
with $c_k$$\geq$0 and $\alpha$=$\sqrt{m/k_B T}$ \cite{pre-I}.

The rate $\lambda'$ defines a decay time scale for each collision event,
$\tau_c$=$1/\lambda'$.
If $\tau_c$ is relatively small (collision effects relax relatively
fast), the memory function associated to the shot noise in (\ref{eq5})
is also local in time.
Then, $\eta(t-t')$$\simeq$$\eta \ \! \delta(t-t')$ and, resorting to
the Markovian approximation \cite{gardiner}, Eq.~(\ref{eq5}) reduces
to
\be
 \ddot{x}(t) = - \eta \dot{x}(t) + F[x(t)] + \delta R(t) ,
 \label{eq6}
\ee
with a total constant friction $\eta$=$\gamma$+$\lambda$ and where
$\langle \delta R(t) \delta R(t') \rangle$=$2m\eta k_B T \delta(t'-t)$.
To estimate the value of $\lambda$, the elementary kinetic theory of
transport in gases \cite{mcquarrie} can be considered, where diffusion
is proportional to the mean free path $\bar{l}$, and the latter is
proportionally inverse to both the density of gas particles and the
effective collision area when a hard-sphere model is assumed.
By means of simple arguments \cite{pre-I}, it is possible to find an
also simple relationship between $\lambda$ and $\theta$, given by
$\lambda$=$(6 \rho \theta / a^2) \sqrt{k_B T/m}$.
Accordingly, increasing $\theta$ (and/or $T$) means an also increase
of the collision frequency.

\begin{figure}[b]
 \includegraphics[width=6cm]{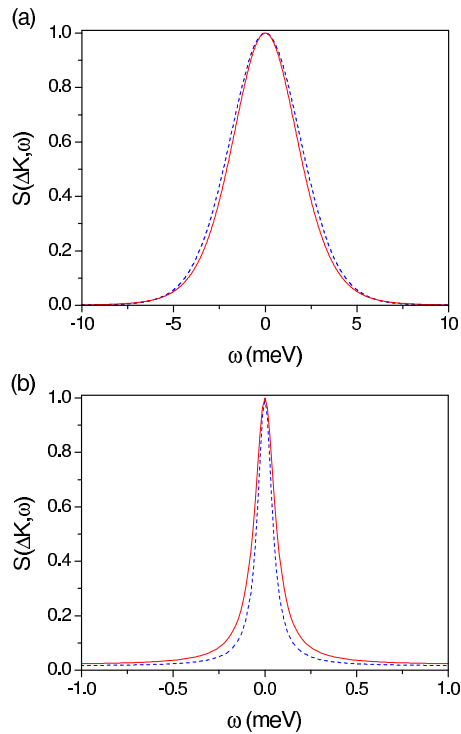}
 \caption{\label{fig1}
  (Color online.) $Q$ peak for Na on Cu(001) at
  $\Delta K$=1.23~\AA$^{-1}$, $T$=200~K, and $\theta_1$=0.028
  (blue dashed line) and $\theta_2$=0.18 (red solid line).
  (a) Flat surface potential ($V$=0) and (b) non separable
  corrugated potential \cite{toennies2}.}
\end{figure}

In Fig.~\ref{fig1} we have plotted 2D calculations for
$S(\Delta {\bf K},\omega)$ with two different types of corrugation:
(a) low corrugation, assuming $V$=0, and (b) a non separable
corrugated potential \cite{toennies2}.
We have considered Na on Cu(001) at $\Delta K$=1.23~\AA$^{-1}$,
surface temperature of 200 K and
two coverages, $\theta_1$=0.028 and $\theta_2$=0.18, as in the
experiment \cite{toennies3}.
After calculating  2D trajectories from Eq.~(\ref{eq6}),
$I(\Delta {\bf K},t)$ is obtained using the rhs of the first
equality in Eq.~(\ref{eq3}) and fitted to a certain analytical function
(see below). Finally, the Fourier transform of that function gives
$S(\Delta {\bf K},\omega)$.
As is clearly seen in Fig.~\ref{fig1}(a), the $Q$ peak gets narrower
as $\theta$ increases for low corrugation.
On the contrary, the presence of a potential leads to the broadening
observed in Fig.~\ref{fig1}(b).
For instance, the full widths at half maximum (FWHM) for three
coverages at $\Delta K$=1.26~\AA$^{-1}$ are: $\Gamma$=120, 190, and
240~$\mu$eV for $\theta$=0.028, 0.106, and 0.18, which are in good
agreement with the experimental ones, $\Gamma$=110, 150, and
270~$\mu$eV, respectively; these widths differ from those found in
Ref.~\onlinecite{toennies3} ($\Gamma$=110, 130, and 160~$\mu$eV,
respectively) by using a dipole-dipole interaction in Langevin
molecular dynamics simulations.

To understand our results we can resort to two (analytical) limit
cases: pure diffusion and anharmonic oscillations.
After assuming $V$=0 in Eq.~(\ref{eq6}), one easily obtains
$\mathcal{C}(t)$=$\langle v_0^2 \rangle e^{- \eta t}$ (where
$\langle v_0^2 \rangle$=$k_B T/m$), which leads to
\be
 I(\Delta {\bf K},t) = \exp \left[- \chi^2
   \left( e^{- \eta t} + \eta t - 1 \right) \right]
 \label{eq7}
\ee
after using Eq.~(\ref{eq4})
with $\chi^2$=$\langle v_0^2 \rangle \Delta K^2 / \eta^2$.
On the other hand, for a damped harmonic oscillator
\cite{JLvega1,risken} we have $\mathcal{C}(t)$=$\langle v_0^2 \rangle
e^{- \eta t/2} \cos (\bar{\omega} t + \delta)/\cos \delta$,
with $\bar{\omega}$=$\sqrt{ \omega_0^2 - (\eta/2)^2}$,
$\delta$=$(\tan)^{-1}(\eta/2\bar{\omega})$, and $\omega_0$
being the harmonic frequency. Again, using Eq.~(\ref{eq4}), we obtain
\begin{equation}
 I(\Delta {\bf K},t) =
  \exp \left\{ - \frac{\chi^2 \eta^2}{\bar{\omega} \omega_0}
 \left[ \cos \delta - e^{-\eta t/2} \cos (\bar{\omega} t - \delta) \right]
  \right\} .
 \label{eq8}
\end{equation}
Unlike (\ref{eq7}), Eq.~(\ref{eq8}) displays an oscillatory
but exponentially damped behavior which does not decay to zero.
Nevertheless, it approaches (\ref{eq7}) in the limit
$\omega_0 \to 0$.
Here we are not interested in the damped harmonic oscillator, but in an
anharmonic one, which arises when the parameters in (\ref{eq8})
are left free.

If deviations from the Gaussian
approximation are not very important, $I(\Delta {\bf K},t)$ will
display features typical of the behaviors described by both (\ref{eq7})
and (\ref{eq8}).
Indeed, even if the Gaussian approximation does not hold, one can still
make use of a working formula in order to extract relevant information
about the process from the numerical results.
Trajectories display both temporary trapping in potential wells
and periods of time where the flight is unbound.
Therefore, assuming a model of {\it running} ($R$) and {\it bound} ($B$)
trajectories, one can consider $\mathcal{C}(t)$=$\alpha \mathcal{C}_R(t)
+ (1 - \alpha) \mathcal{C}_B(t)$, where $\mathcal{C}_R$ and
$\mathcal{C}_B$ correspond to the velocity autocorrelation functions
for a flat surface and a damped anharmonic oscillator, respectively.
$I(\Delta {\bf K},t)$ can be then written as
\be
 I(\Delta {\bf K}, t) \approx [I_R (\Delta {\bf K}, t)]^\alpha
  [I_B (\Delta {\bf K}, t)]^{1 - \alpha} ,
 \label{eq9}
\ee
where $I_R$ and $I_B$ are given by Eqs.~(\ref{eq7}) and (\ref{eq8}),
respectively.
This expression, with free parameters, is used as our working formula
and allows for a distinction between the contributions arising from
the unbound or diffusive motion and the bound or vibrational one.
\begin{figure}[b]
 \includegraphics[width=6.2cm]{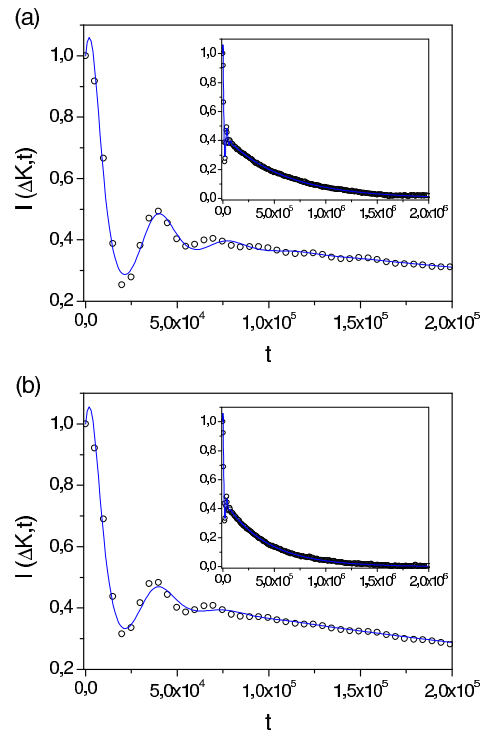}
 \caption{\label{fig2}
  (Color online.) $I(\Delta {\bf K}, t)$ for Na on Cu(001) at
  $\Delta K = 1.23$~\AA$^{-1}$, $T$=200~K, and
  (a) $\theta_1$=0.028 and (b) $\theta_2$=0.18.
  Open circles indicate the numerical values obtained from the
  simulation and solid lines are the numerical fitting using
  Eq.~(\ref{eq9}).}
\end{figure}
As seen in Fig.~\ref{fig2}, Eq.~(\ref{eq9}) fits fairly well the
Langevin numerical results obtained for the corrugated surface
potential [see Fig.~\ref{fig1}(b)] and both coverages.
The fitting parameters compared with the nominal values are given in
Table~\ref{tab1}; though fitted and nominal values are different,
their order of magnitude and trend are correct, regarding such
differences to deviations from the Gaussian approximation.
More importantly is, however, the resulting value for $\alpha$:
$\alpha_1$=0.0294 for $\theta_1$, and $\alpha_2$=0.0410 for
$\theta_2$.
From these values, it is clear that the contribution to
$I(\Delta {\bf K}, t)$ primarily comes from bound motions.
Note from (\ref{eq7}) that
running trajectories lead to a relative much faster decay of
$I(\Delta {\bf K}, t)$ than the bound motion [see Eq.~(\ref{eq8})],
which delays such a decay.
This is so because the bound motion keeps correlations for times longer
than the diffusive one, since the latter provokes a fast {\it dephasing}
of the (correlation) oscillating terms that appear in the rhs of the
first equality in Eq.~(\ref{eq3}).
In this way, this explains, first, that $S(\Delta {\bf K}, \omega)$
is about two orders of magnitude broader in the flat case than in the
corrugated one (see Fig.~\ref{fig1}).
And, second, since $\alpha_2$$>$$\alpha_1$ there is a slightly larger
fraction of running trajectories for $\theta_2$, which leads to an
also slightly faster decay of $I(\Delta {\bf K}, t)$, and therefore
to observe broadening in $S(\Delta {\bf K}, \omega)$ with increasing
$\theta$.
It is also worth stressing that, with increasing coverage, diffusion
decreases (according to Einstein's relation, it goes like $\eta^{-1}$)
and the running trajectories display short flights in small surface
regions.

\begin{table}
 \begin{tabular}{c p{.1cm} c p{.1cm} c p{.1cm} c p{.1cm} c}
  \hline\hline
   \ coverage         & & $\omega$ (a.u.)   & &  $\eta$ (a.u.)  & & $\delta$ & & $\chi$  \\ \hline
   \ $\theta_1^{\rm nom}$ & & 2.20$\times$10$^{-4}$  & & 2.54$\times$10$^{-5}$ & &  0.058 & &  3.15 \\
   \ $\theta_1^{\rm fit}$ & & 1.39$\times$10$^{-4}$  & & 3.52$\times$10$^{-5}$ & &  0.184\ \ \  & &  4.64 \\
  \hline
   \ $\theta_2^{\rm nom}$ & & 2.19$\times$10$^{-4}$  & & 4.69$\times$10$^{-5}$ & &  0.107 & &  1.71 \\
   \ $\theta_2^{\rm fit}$ & & 1.37$\times$10$^{-4}$\ \ \ & & 4.28$\times$10$^{-5}$ & &  0.276\ \ \ \   & &  2.82 \\
  \hline\hline
 \end{tabular}
 \caption{\label{tab1} Nominal and fitted values involved in the
  calculation of $I(\Delta {\bf K}, t)$ for $\theta_1$=0.028 and
  $\theta_2$=0.18.}
\end{table}

To conclude, the simple stochastic model together with the theoretical
analysis presented in this work shows that the experimental broadening
observed in the $Q$ peak with increasing $\theta$ (from low to
intermediate regimes) is related to the bound motion undergone by the
adsorbates inside the potential wells, and not only to purely diffusive
motion or to the particular form assumed for adsorbate-adsorbate
interactions.
Moreover, regardless the adsorbate-substrate interaction potential,
in our model the line shape of $S(\Delta {\bf K},\omega)$ is also
related to $\eta$, which gathers the thermal effects caused by the
surface ($\gamma$) and the collisions among adsorbates ($\lambda$).
Since $\gamma$ is assumed to be constant, the broadening is thus
directly related to an increasing friction $\lambda$ as $\theta$
increases.
The idea of replacing the dipole-dipole interaction by a shot noise
is crucial because, for long time processes with a high number of
collisions, the statistical limit seems to wipe out any trace of
the true interaction potential. As happens with the models proposed in
the literature \cite{toennies2,toennies3,benedek}, ours gives a
smaller diffusion than expected with increasing $\theta$, which could
be related to the fact that $\theta$ might also influence the type of
adsorbate-substrate interaction, and therefore modify parameters
such as the activation barrier or the surface friction in our model.
Thus, in Ref.~\onlinecite{benedek}, the Lau-Kohn long range
interaction via intrinsic surface states is proposed to use a
$\theta$-dependent surface electronic density modifying the
corrugated potential and to explain the remarkable increasing
of the $T$ mode frequency with coverage at 50~K which a dipole-dipole
or substrate mediated coupling cannot explain it.
This increasing is also observed at 200~K and parallel momentum
transfer of $- 2$~\AA$^{-1}$ to be 11$\%$, which is similar to our
value (around 7$\%$) when going from $\theta$=0.028 to 0.106 at
2.10~\AA$^{-1}$.
Although further investigation at microscopic level and calculations
from first principles are needed, our simple stochastic model at low
coverages and surface temperatures is also able to provide a
complementary view of diffusion and low-frequency vibrational motions,
described by the peaks around or near zero energy transfers in the
scattering law.

This work has been supported in part by DGCYT (Spain) under project
FIS2004-02461.
R.M.-C.\ thanks the University of Bochum for support from the
Deutsche Forschungsgemeinschaft, SFB 558, for a predoctoral contract.
J.L.V.\ and A.S.\ Sanz would like to thank the Mi\-nis\-te\-rio de
Educaci\'on y Ciencia (Spain) for a predoctoral grant and a ``Juan
de la Cierva'' Contract, respectively.

%%%%%%%%%%%%%%%%%%%%%%%%%%%%%%%%%%%%%%%%%%%%%%%%%%%%%%%%%%%%%%%%%%%%%%%

\end{document}